\input harvmac
\input epsf.tex       
$\,$ 
\overfullrule=0pt 
\vskip-2cm
\Title{\vbox{\baselineskip12pt\hbox{\phantom{lll}}\hbox{}}}
{\vbox{\centerline{Maximal Non-Abelian  
Gauges and  Topology} 
\medskip 
\centerline { of Gauge Orbit Space}}} 
\medskip  
\centerline{\bf M. Asorey} 
\bigskip\centerline{{ Departamento de F\'{\i}sica Te\'orica.
Facultad de Ciencias}} \centerline{Universidad de Zaragoza.  50009
Zaragoza. Spain}

\font\frak=eufm10 scaled\magstep1

 2 

scaled\magstep3 
 
\font\tenfrak=eufm10

\def\sevenpoint{\def\rm{\fam0\sevenrm}
\textfont0=\sevenrm \scriptfont0=\fiverm \scriptscriptfont0=\fiverm
\textfont1=\seveni  \scriptfont1=\fivei  \scriptscriptfont1=\fivei
\textfont2=\sevensy \scriptfont2=\fivesy \scriptscriptfont2=\fivesy
\textfont\itfam=\tenit \def\it{\fam\itfam\tenit}
\textfont\bffam=\tenbf \def\bf{\fam\bffam\sevenbf} \rm}

\def\goth #1{\hbox{{\frak #1}}} 

\def\smallgoth #1{\hbox{{\tenfrak #1}}}

\def\frac#1#2{{#1\over #2}}

 
\def\bbig R{{\hbox{{\bbigfield R}}}}

%
\def\inbar{\,\vrule height1.5ex width.4pt depth0pt}
\def\IB{\relax{\rm I\kern-.18em B}}
\def\IC{\relax\hbox{$\inbar\kern-.3em{\rm C}$}} 
\def\IIC{\relax{\rm I\kern-.18em C}} 
\def\ID{\relax{\rm I\kern-.18em D}}
\def\IE{\relax{\rm I\kern-.18em E}} 
\def\IF{\relax{\rm I\kern-.18emF}} 
\def\IG{\relax\hbox{$\inbar\kern-.3em{\rm G}$}}
\def\IH{\relax{\rm I\kern-.18em H}} 
\def\II{\relax{\rm I\kern-.18em I}} 
\def\IK{\relax{\rm I\kern-.18em K}} 
\def\IL{\relax{\rm I\kern-.18em L}} 
\def\IM{\relax{\rm I\kern-.18em M}} 
\def\IN{\relax{\rm I\kern-.18em N}}
\def\IO{\relax\hbox{$\inbar\kern-.3em{\rm O}$}} 
\def\IP{\relax{\rm I\kern-.18em P}} 
\def\IQ{\relax\hbox{$\inbar\kern-.3em{\rm Q}$}}
\def\IR{\relax{\rm I\kern-.18em R}} 
\font\cmss=cmss10
\font\cmsss=cmss10 at 10truept
\def\ic{\relax{\rm \kern.2emI\kern-.4em C}} 
\def\IZ{\relax\ifmmode\mathchoice {\hbox{\cmss
Z\kern-.4em Z}}{\hbox{\cmss Z\kern-.4em Z}} {\lower.9pt\hbox{\cmsss
Z\kern-.36em Z}} {\lower1.2pt\hbox{\cmsss Z\kern-.36em
Z}}\else{\cmss Z\kern-.4em Z} \fi} 
\def\IGa{\relax\hbox{${\rm I}\kern-.18em\Gamma$}} 
\def\IPi{\relax\hbox{${\rm I}\kern-.18em\Pi$}} 
\def\ITh{\relax\hbox{$\inbar\kern-.3em\Theta$}}
\def\IOm{\relax\hbox{$\inbar\kern-3.00pt\Omega$}}
\def\bupnn{\relax\hbox{ $\cdots
\kern-.3cm\hbox{\raise 1ex\hbox{$^{n}$}}$}} 
\def\buppn{\relax\hbox{ $\cdots
\kern-.5cm\hbox{\raise 1ex\hbox{$^{d-n}$}}$}} 
\def\buppd{\relax\hbox{ $\cdots \kern-.5cm\hbox{\raise
1ex\hbox{$^{d-2}$}}$}} 
\def\buppu{\relax\hbox{ $\cdots
\kern-.5cm\hbox{\raise 1ex\hbox{$^{d-1}$}}$}} 
\def\cite{{\ref\uns}}
\def\CC{{\cal C}}
 
\def\CQ{{\cal Q}} 
\def\CP{{\cal P}} 

\def\CM{{\cal M}}

\def\CH{{\cal H}}

\def\CG{{\cal G}}

\def\cm{{\smallgoth m}} 
\def\Cm{S{\smallgoth M}} 
\def\su{{\smallgoth s u}} 
\def\h{{\goth h}}
\def\j{{\goth j}} 
\def\n{\hbox{\cmsss (N)}}
\def\nn{\hbox{\cmsss N}}

\def\hn{{\goth h}_{\nn}}
 
\def\tr{\hbox{{\rm tr}}}

\def\Maps{\hbox{{\rm Maps}$_0\,$}} 
\catcode`\@=11
\font\tenbmi=cmmib10 
\skewchar\tenbmi='177 
\font\sevenbmi=cmmib10 at
7pt \skewchar\sevenbmi='177 
\font\fivebmi=cmmib10 at 5pt
\skewchar\fivebmi='177 
\newfam\bmfam  

\textfont\bmfam=\tenbmi 
\scriptfont\bmfam=\sevenbmi
\scriptscriptfont\bmfam=\fivebmi 
\def\back{{{\raise.4em\hbox{$\, _\backslash\,$}}}} 
\catcode`\@=12 %

\baselineskip=16pt plus 2pt minus 1pt  
\bigskip  
   
\centerline{\bf Abstract} 
\bigskip 
  
We introduce two
maximal non-abelian gauge fixing conditions on the space of gauge 
orbits $\CM$ for gauge theories over spaces with dimensions $d\leq 3$. 
The gauge fixings are {\it complete} in the sense that describe an open
dense set $\CM_0$ of the space of gauge orbits $\CM$ and  select
one and only one gauge field per gauge orbit in $\CM_0$. There
are not Gribov copies or ambiguities in these gauges. $\CM_0$ is a
contractible manifold with trivial topology. The set of gauge orbits
which are not described by the gauge conditions $\CM{\back}\CM_0$ is
the boundary of $\CM_0$ and encodes all non-trivial topological 
properties of the space of gauge orbits. The gauge fields
configurations of this boundary $\CM{\back} \CM_0$ can be
explicitly identified with non-abelian monopoles and they are
shown to play a very relevant role in the non-perturbative 
behaviour of gauge theories in one, two and three space dimensions.
It is conjectured that their role is also crucial for quark
confinement in 3+1 dimensional gauge  theories.

 \overfullrule=0pt  
\hyphenation{systems}  
\bigskip\vfill
\noindent\baselineskip=16pt plus 2pt minus 1pt 
\Date{ }  
\newsec{Introduction}

One of the prominent features of non-abelian gauge theories is the
highly non-trivial geometric and topological structure of the space
of physically relevant classical gauge field configurations. 
Because of gauge invariance this space is the  gauge orbits space 
$\CM=\CA/\CG$, i.e., the space of gauge fields $\CA$ modulo the
group of gauge transformations $\CG$. The first evidence of the
non-trivial structure of $\CM$ came from Gribov's observation on the
existence of
ambiguities and possible incompleteness of Coulomb and Landau gauges
in Yang-Mills theories \ref\gribov{V. N. Gribov, Nucl. Phys. {\bf B
139}  (1978) 1}. The impossibility of a global gauge fixing was shown
to be  a consequence of the non-trivial topological structure of the
space  of gauge orbits  \ref\singer{I. M. Singer, Comm. Math. Phys.
{\bf 60} (1978) 7}.  However, the need of efficient gauge fixing is
a requirement for analytic approaches to the dynamical behaviour of 
the theory both in the asymptotically free ultraviolet regime 
(perturbative) and in the confining infrared regime 
(non-perturbative).

Landau gauge and its $\alpha$--gauge generalizations played a leading
role in the development of the perturbative renormalization program of 
quantum gauge theories  because of its covariant linear and local
characteristics \ref\fp{L. Faddeev and V.  Popov, Phys.
Lett. {\bf B 25} (1967) 29}. Coulomb gauge  was very useful in the
formulation of the Hamiltonian approach  \ref\sch{J. Schwinger,
Phys. Rev. {\bf D 127}(1962)  324}.        

The Gribov observation, however, points out the existence of a 
possible  problem by under/over--counting the quantum fluctuations of
classical gauge fields on these gauges.  The problem does not affect
 perturbative calculations because the Gribov horizon and the
Gribov copies give contributions of order $e^{-1/g^2}$ \ref\aph{
M. Asorey and F. Falceto, Ann. Phys. N.Y.  {\bf 196} (1989) 209}.
There are, however, perturbative effects like anomalies which are
very sensitive to the global structure  of the space of gauge
orbits. In fact, anomalies provide  the most direct evidence
that the non-trivial topological structure of $\CM$ turns out to be
relevant for the quantum physics. The reason why anomalies which
already appear in perturbation theory can unveil the non-trivial
topological structure of the gauge orbit space is due to
universality and non-renormalization  theorems. They essentially
stablish that the anomaly structure is stable for all energy scales.
In the ultraviolet regime they are determined by perturbative
methods, but in the infrared regime they are extremely connected
with the cohomology  of the space of gauge orbits $\CM$. Whether the
topology of the orbit space is relevant or not for other
non--perturbative effects like confinement requires a deeper study
incorporating the contributions of gauge field configurations
affected by the Gribov problem. 

To circumvent this problem and 
include the non-perturbative effects associated to those
configurations there are  three alternatives:

i) Restrict the analysis to  subsets of gauge fields satisfying
gauge conditions which are free of ambiguities (no overcounting
gauge orbits). For instance,  consider several Coulomb/Landau gauges
around different background fields. Then, consider a complete
covering of the orbit space  by means of those   subsets
 and use a partition of unity to get the right contributions to the
quantum fluctuations.  Obviously, the procedure is rather
cumbersome, but consistent. The BRST symmetry is also consistent in
the overlap of those subsets of gauge fields satisfying   different
gauge conditions which means that it  can be globally defined over
the whole space of gauge orbits and guarantees the consistency of
the whole procedure \aph.

ii) Prove that any gauge orbit is described by few fields in a
particular gauge condition and that the sum over all  Gribov copies
gives the right contribution up to a global factor associated to the
volume of overcounted copies. Although a gauge condition with  this
property is hard to implement  in  the continuum, which is absolutely 
necessary to define a consistent quantum theory, in the lattice approach 
it works for some gauge conditions 
\ref\brand{B. Sharpe, J. Math. Phys. {\bf 25} (1984) 3324 }.

iii) Find a special set of gauge fields satisfying a single gauge
condition such that their gauge orbits unambiguously describe an open
dense subset of the space of gauge  orbits. Then, the fluctuations
associated to those gauge fields are enough to describe the quantum
effects of the theory. In the Hamiltonian formalism the role of the
remaining orbits re-emerge as boundary conditions on physical
quantum states.

These ways of solving the Gribov  problem   can be illustrated with
the following example. Let us consider the quantization of classical
system on $\IR^3$ with a kinetic term independent of the radial
degree of freedom. The effective configuration space is
 a 2-dimensional sphere $S^2$. This space  can
be considered as the space of the orbits of the physical space
$\IR^3-\{x_0\}$ under dilations, $S^2= \IR^3/\IR^+$.  The first
method can be implemented by the choice of four planes intersecting 
in a cube around the origin of $\IR^3$. The parametrizations of  the
dilation group orbits given by these four charts cover the whole
$S^2$ sphere. The use of spherical coordinates provides a
parametrization of the orbits in terms of angular variables by a
simple chart that only excludes the two azimuthal points  of $S^2$.
This provides an example of gauge fixing of the third  type. There
are many other ways of parametrizing the 2-sphere 
not always directly connected with dilation orbits of $\IR^3$ (e.g. by
stereographic projection coordinates).

It is obvious that  the third type of gauge  is the most
 economic for the description of the dynamics.  The aim of this
paper is to find a gauge condition of this type for Yang-Mills
theory. 

In Landau gauge it has been proved that it is possible to
find a subset of configurations (fundamental domain) which
parametrize a dense set of orbits with respect to the $L^2$--norm of
the space of gauge field  configurations $\CA$ \ref\sem{M.A.
Semenov-Tyan-Shanskii and  V.A. Franke, Pubs. LOMI Seminar {\bf 120} 
(1982) 159; English translation:
 J. Sov. Math. {\bf 34} (1986) 1999}, although the construction of
the fundamental domain is not achieved in a very explicit way.
 However, the main problem is that the $L^2$--norm is not
 relevant for the measure of  quantum fluctuations. The leading
contributions to the functional integral come from more singular 
configurations and when  we consider a continuum regularization
\ref\am{M. Asorey and P.K. Mitter,  Commun.
Math. Phys. {\bf 80} (1981) 43 }  the relevant contributions need
to be smoother than those described by the $L^2$--norm to
guarantee that  the orbit space is a smooth manifold
and that the functional measure is well defined globally on $\CM$
\am\ref\af{M. Asorey and F. Falceto,  Nucl. Phys. {\bf B 327} (1989)
427}. There not exists a  generalization of the 
Semenov-Tyan-Shanskii-Franke result to guarantee the existence of a 
similar domain in Landau gauge  for  smooth ($C^\infty$) 
or  Sobolev (with $k>1+d/2$)
gauge fields  for spaces of dimension $d>1$ 
\ref\as{M. Asorey, Unpublished (1987)}. 

In this paper we introduce a different gauge fixing method which leads
to a complete parametrization of a dense set of gauge orbits in the
space of gauge fields not only in the $L^2$--norm but also in the
$C^\infty$-smooth and Sobolev  topologies of gauge fields for spaces with
dimensions lower than four. The special configurations whose orbits
are at the boundary of this domain have a peculiar dynamical
behaviour which can be related to  non-perturbative effects of
the theory. Another relevant feature of this novel gauge fixing is
that is very explicit and the fundamental domain can be identified
without ambiguities.

Another type of gauge conditions which has also been 
extensively considered in the recent literature are the abelian gauges
introduced by 't Hooft  some years ago \ref\th{G.~'t~Hooft,
Nucl.\ Phys.\ {\bf B190} (1981) 455.}. They allow
the  identification of some classical configurations with magnetic
monopoles and numerical simulations suggest that those configurations
carrying a magnetic charge play a leading role in   confinement
mechanisms. Abelian gauge conditions involve partial gauge fixing:
the  gauge is fixed in the non-abelian modes and  abelian modes are
left without  gauge fixing which allows the description of abelian 
magnetic monopoles.  These gauge conditions are essentially  defined 
by means of an auxiliary gauge covariant functional $\Phi(A)$. The 
dependence on the choice of this functional makes unclear the intrinsic 
physical role of the configurations carrying magnetic charges. The most 
popular abelian gauge, the maximal abelian gauge, which is well defined 
on the lattice formulation  \ref\mab{A.S.~Kronfeld,
M.L.~Laursen, G.~Schierholz and U.-J.~Wiese, Phys.\ Lett.\ {\bf B198}
(1987) 516}  is not necessarily complete in continuum space-times\th.
Another special gauge of this type is the Laplacian abelian gauge
where the functional $\Phi(A)$ is chosen as the lowest 
eigenfunction of the covariant laplacian operator $\Delta_A$
\ref\arjan{ A.J. van der Sijs, Nucl.\ Phys.~B (Proc.\ Suppl.)\ {\bf
53} (1997) 535}. This is not uniquely defined for  gauge field
configurations with degenerated ground states of $\Delta_A$.

The Maximal Non-Abelian Gauges that we introduce in this paper shares
with abelian  gauges some features like the  association of magnetic
charges to gauge configurations but has the advantage of being
intrinsically defined. The magnetic charge of a gauge configuration
is not a gauge fixing artifact like in the abelian gauges. On the
other hand the gauge conditions are uniquely defined without ambiguities 
and are complete. Although the domain of the orbit space where the gauges
are  defined is contractible and thus topologically trivial, the
non-trivial nature of the whole orbit  space is recovered when we
add the gauge orbits sitting at the boundary of the fundamental
domain.

The outline of the paper is as follows. We review in Section 2 the
main topological properties of the orbit space of gauge fields. In
Sections 3 and 5 we analyze the structure of the orbit spaces of one
and two dimensional gauge fields which turn out to be the essential
ingredients for the definition of the maximal non-abelian gauges.   
These are explicitely introduced and analyzed in sections 4 and 6. 
The relation  of non-abelian monopoles  with gauge configurations
which lie beyond the boundary of the fundamental domain of maximal
non-abelian gauges is analyzed in section 7, where it is also
discussed the relevance of those configurations for different
non-perturbative effects. We conclude in section 8 with a summary of
the main results.

\newsec{Topological structure of the gauge orbit space $\CM$}

Let us consider SU(N) gauge fields defined on a d-dimensional sphere
$S^d$. The infinite volume case requires a separate discussion
and  will be considered below. The action of the group of gauge
transformations $\CG$ in the space of gauge fields $\CA$ is not free
even if we mod out by the center of the  group  $Z_{\nn}$ which
obviously does not act  on  $\CA$. This problem is due to the
existence of reducible gauge connections with smaller holonomy groups 
SU(N'), $N'<N$. Their orbits have a larger isotopy group which generates
singularities in the orbit space. There are two ways of
circumventing this problem. The first option consists in  do not 
consider those singular gauge orbits at all. Indeed,  irreducible
gauge fields  define an open dense subspace of $\CA$ and for any
Borel measure on $\CM$ the reducible orbits will have zero measure.
However, important configurations belong to the class of reducible
gauge fields (e.g. classical vacua) and it is not very 
reasonable to exclude those fields from the physical
configuration space. We shall consider an alternative option which
proceeds by  considering all gauge fields but modding out only by
the group of pointed gauge transformations $\CG_0$, that is the
group of gauge transformations with reduce to identity  for a fixed
given point of $S^d$. This group has no center and  acts freely on
$\CA$. The corresponding quotient space $\CM=\CA/\CG_0$ is also a
smooth manifold \foot{We  consider (infinitely) differentiable
gauge fields and gauge transformations, but similar results will
hold for gauge fields in a high enough Sobolev class, $k>{d\over 2}+1$}. 
The only problem of this approach based on the  pointed orbit space is that
the full gauge group is a symmetry of the dynamics which implies that
there is  residual global symmetry which  leads  to the existence of
zero modes in the propagators  in perturbation theory. The problem is
solved by the introduction of collective coordinates. From a
non-perturbative point of view the analysis is completely consistent
because the contribution of the fields associated to the residual
symmetry is always finite and controlled by the volume of the gauge
group SU(N).

The fact that $\CA$ is an affine space implies that its  topology is
trivial. However, the gauge group $\CG_0$ exhibits a rather
non-trivial topological structure. Since $\CA(\CM,\CG_0)$ has a
principal bundle structure and $\CA$ is homotopically trivial the
homotopy groups of the orbit space are given by
$$\pi_n(\CM)=\pi_{n-1}(\CG_0)=\pi_{n+d-1}(SU(N)).$$ In particular
this means that the first non-trivial homotopy groups in two, three
and four dimensions are $\pi_2(\CM^{S^2})=\IZ$, $\pi_1(\CM^{S^3})=\IZ$ and 
$\pi_0(\CM^{S^4})=\IZ$, respectively. In the last case, the different
connected components of the orbit space correspond to classes of
gauge fields defined on different principal bundles parametrized by
the second Chern class $c_2$. This property of four dimensional 
space-times also holds for higher dimensional space-times, i.e.
$\pi_0(\CM)=\IZ$. Of course every  connected component of
$\CM$  also has non-trivial higher homotopy groups.  
 
Because of the non-trivial structure of the bundle  $\CA(\CM,\CG_0)$
there are not   continuous global sections, i.e. it is impossible to
fix a  continuous global gauge \singer.

For the same reasons the cohomology groups  of $\CM$ are
non-trivial. They are given by $$ H^n(\CM,\IR)= \IZ^{r_n},$$ where
$r_n$ is the coefficient of the $t^n$ term in the series 
\eqn\poinc{\eqalign{\CP(t)=&(1-t^2)^{-1} (1-t^4)^{-1} \cdots
(1-t^{2N-d})^{-1}  \qquad \hbox{$\,$\rm for even d}\cr \CP(t)=&
(1+t) (1+t^3) \cdots (1+t^{2N-d}) \qquad\qquad \qquad\hbox{\rm for
odd d}}} 
Many of those cohomology groups are associated with gauge
anomalies of quantum field theories with fermion fields.

This highly non-trivial topological structure of the orbit space is
accompanied of a non-trivial Riemannian geometric structure  induced
by the  dynamics of gauge field theories \ref\BV{O. Babelon  
and C. Viallet, {Comm. Math. Phys.}
{\bf 81} (1981) 515}. 
One of the most relevant consequences of this structure is
the absence of a standard particle interpretation of the energy
spectrum. In pure scalar (or fermionic) field theories the
configuration space of classical fields is a linear space $\CQ$ and
the quantum states are functionals of the space $L^2(\CQ,\mu)$ of
square integrable functionals with respect to a probability measure
$\mu$  of $\CQ$  associated to the quantum vacuum. One relevant
subspace of $L^2(\CQ)$ is the space of linear functionals $\CQ^\ast$
which can be identified with the one particle states of the quantum
theory. It is obvious that $L^2(\CQ,\mu)$ can be identified with 
Fock space $\oplus_{n=0}^\infty\CH^n$ associated to $\CH=\CQ^\ast$. 
This identification makes
natural the particle interpretation of the energy spectrum. In the 
case of gauge theories the physical states are functionals on the
space $L^2(\CM,\mu)$ and  because of the curved nature of $\CM$
there is no analogue of the  subspace of one particle states
$\CQ^\ast$. This does not exclude  a particle interpretation of the
spectrum but makes it less evident to hold for all energy scales.

{\it Remark:} In infinite volumes the gauge fixing problem becomes
more subtle. Naively speaking there is no gauge fixing problem for
$C^\infty$ --smooth gauge fields. The reason being that the group of
smooth  $C^\infty$ gauge transformations is a contractible manifold
with trivial topology. In that case Singer's proof of topological
obstruction to the existence of a global gauge fixing fails. Indeed,
any given $C^\infty$-smooth  pointed gauge transformation over $\IR^d$,
 $\phi: \IR^d\to SU(N)$, with $\phi(0)=\II$  can be
homotopically contracted to the trivial transformation $\phi(x)=\II$
by the map 
\eqn\contr{\phi(t,x)=\phi({\rm e}^{(1-1/t)}x)\qquad
t\in [0,1],} 
which obviously is continuous and smooth and interpolates between 
$\phi(1,x)=\phi(x)$ and $\phi(0,x)=\II$.

    However, in  the quantum theory the relevant fields 
satisfy some regularity conditions at infinity. For instance, in 
the ultraviolet regularized theory the relevant fields belong to
an appropriate  Sobolev class which implies that they satisfy
specific boundary conditions. This makes the gauge fixing
problem relevant for the quantum field theory. In order to not
prejudge the infinity volume behaviour we can choose an $S^{d-1}$
fibration of $R^d$ given by radial spheres centered at the origin of
$R^d$. The gauge fields are parametrized by a family of gauge fields
defined on the different $S^{d-1}$ radial shells of $\IR^d$ and a
family of (radial) Higgs fields defined on the same spherical
shells. Quantum dynamics imposes additional conditions on the behaviour
of the fields for infinite radius. Anyhow, this shows that the 
study of gauge fields on  spheres is also relevant for the description 
of the infinite volume limit.

\newsec{Two Dimensional gauge fields}

In two dimensions any principal SU(N) bundle $P(S^2,SU(N))$
is trivial $P=S^2\times SU(N)$. The two-dimensional sphere $S^2$ 
also  has a  natural complex structure. This makes possible to
identify the space of SU(N) gauge fields on $S^2$  with the space of
SL(N,$\IC$) holomorphic bundles on  the trivial vector bundle
$E(S^2,\IC^{\nn})$ with $E=S^2\times\IC^{\nn}$ \ref\ab{ M. Atiyah and
R. Bott,  Philos. Trans. Roy. Soc. London  {\bf A308}  (1982) 523}.

A SL(N,$\IC$) holomorphic bundle is  a vector bundle structure
 with holomorphic transition functions, i.e. the transition
functions of a holomorphic bundle $g_{ij}(x)\in SL(N,\IC)$ must
satisfy, besides the compatibility conditions  $$ \eqalign{ 
g_{ii}(x)=\II_{\nn} \qquad &{\hbox{\rm  for}}\  x\in U_i\cr
g_{ij}(x)\  g_{jk}(x)\   g_{ki}(x) =\II_{\nn} \quad &{\hbox{\rm
for}}\   x\in U_i\cap U_j\cap U_k,\cr } $$
 the holomorphic condition $$ \partial_{\bar z} g_{ij}(x)=0 \qquad
{\hbox{\rm  for}}\ x\in  U_i\cap U_j.$$

Given a SL(N,$\IC$)
holomorphic bundle in $E$  there exists a unique SU(N) gauge
field $A\in \CA$ whose covariant derivative operator $D=d_A$ 
satisfies  $D_{\bar z} \sigma=0$ for any local holomorphic section
$\sigma$ of $E$. We consider the trivial hermitean structure of
$E=S^2\times\IC^N$ induced by the scalar product of $\IC^N$.

In local coordinates $D_{\bar z}=\partial_{\bar z} + { h}^{-1}
\partial_{\bar z} { h}$,  where ${h}:U\to SL(N,\IC)$ are the
coordinates of a given local  holomorphic frame in 
$E=S^2\times\IC^N$. Notice  that the local expression does not
depend on  the choice of such a  frame and only depends on the
SL(N,$\IC$) holomorphic bundle structure.

Conversely, given a SU(N) gauge field $A$  there exists a unique 
SL(N,$\IC$) holomorphic bundle structure on $E$ whose associated
gauge field is $A$. This follows from the fact that the local
sections $\sigma$ satisfying the condition $D_{\bar z} \sigma=0$
define a  SL(N,$\IC$) holomorphic bundle structure on $E$. 
It can be shown that the
correspondence between SU(N) gauge fields and SL(N,$\IC$)
holomorphic bundle structures is one-to-one in two dimensions. 

In four dimensions there is a similar correspondence, but in such a
case the   corresponding gauge fields associated to   holomorphic
bundles must be  selfdual. In two dimensions there is no constraint
on the associated unitary connections.

The characterization of 2-dimensional gauge fields in terms of
holomorphic bundles has been very useful for  the resolution of
various field theories like the O(3) sigma model
\ref\WPol{A.M. Polyakov and P.B. Wiegmann,  Phys. Lett. {\bf B 141} 
(1984) 223}, Wess-Zumino-Witten theory \ref\gk{K. Gawed\c cki 
and A. Kupiainen, { Phys. Lett.} 
{\bf B 215} (1988) 119; Nucl. Phys. {\bf B320} (1989) 649},
Chern-Simons theories \ref\gkk{
K. Gawed\c cki and A. Kupiainen, Commun. Math. Phys. {\bf 135} 
(1991) 531} and more recently in pure
Yang-Mills theory in three-dimensions 
\ref\nakab{V.N. Nair and D. Karabali, Nucl. Phys. {\bf B464} (1996) 135}.

Holomorphic bundles which are related by a linear homomorphism of
$E$ are said to be equivalent. In terms of the gauge field
representation this induces the following equivalence relation
\eqn\gtr{ A^h=h^{-1}
A_{\bar z} h+ h^{-1}\partial_{\bar z} h,}
given by the action of the group of linear
complex automorphisms $\CG_{\ic} ={\rm Maps}(S^2,SL(N,\IC))$ on the space
of gauge fields $\CA$. $\CG_{\ic}$ is twice larger than the group
 of ordinary gauge transformations $\CG={\rm Maps}(S^2,SU(N))$ because
the automorphisms of $\CG_{\ic}$ correspond to complex gauge 
transformations, i.e.
$\CG_{\ic} \approx\CG\times\CG$. For such  reasons the orbit
space  of this larger symmetry group is smaller than that of $\CG$.
In fact, since the complex gauge transformation involves the same
 number of local degrees of freedom that the two-dimensional gauge
fields (two $\su\n$--valued scalar fields), this
orbit space is expected to be a finite dimensional topological
space. The only problem is that this orbit space $\cm$ has
an stratified structure.

The different isomorphism classes of SL(N,$\IC$) holomorphic bundle 
structures on $S^2$ were classified by Grothendieck and
turn out to define a discrete moduli space. They are given by
the following transition functions connecting two
holomorphic patches defined on the north and south
hemispheres
\eqn\ttr{g_{ii}(x)=
\pmatrix{z^{n_1}&0&\cdots& 0\cr 0&z^{n_2}&\cdots& 0\cr
\cdots&\cdots& \cdots &\cdots\cr 0&0&\cdots & z^{n_N}\cr},} where
$n_i\in \IZ$ for all $i=1,\cdots,N$ 
 and $n_1\geq n_2\geq\cdots\geq n_N$ with $n_1+n_2+\cdots +n_N=0$.
  The fact that the exponents  $n_i$ are integers suggests
that the corresponding gauge fields can carry a non-trivial
non-abelian  monopole structure as we will stress in section 7. The
holomorphic bundle structures associated to non-trivial integers are
called unstable holomorphic bundles according to Mumford's 
classification. The  main difference between the trivial class, where
all $n_i=0$, and the other classes of holomorphic structures which
correspond to unstable  bundles is that in the first case the
transition function \ttr\ is proportional to the identity, i.e. only
one chart is required to describe the corresponding holomorphic
bundle. Thus, all  bundles of this type define a subset $\CC_0$ of
the space of all holomorphic bundles which is
 in one to one correspondence with the space of connections
$\CA_0$ of the form 
\eqn\den{A_{\bar z}=h^{-1}\partial_{\bar z}
h, \qquad h:S^2 \to SL(N,\IC).} 
It is well known that $\CC_0$ is
dense on $\CC$, thus, the space $\CA_0$  of gauge fields of the
type \den\ is also dense in the space of all gauge fields $\CA$. In
consequence, the space of the ordinary gauge orbits of the fields in 
$\CA_0$ define a
dense submanifold $\CM_0$ of the orbit space $\CM$. In fact, the
boundary of this space, $\CM_\ast=\CM\back\CM_0$ which correspond to 
 gauge fields associated to unstable bundles which cannot be
globally written as \den\ is a submanifold of $\CM$ which has
codimension  two \ref\unst{M. Asorey, F. Falceto and G. Luz\'on, 
Contemp. Math. {\bf 219}
(1998)1}\foot{ We shall consider  two types of regularity
conditions on the gauge fields. Either gauge fields in a  Sobolev
class $k>1 + d/2 $ or simply $C^\infty$-smooth gauge fields. 
Although the results hold for more general regularity 
conditions}.

Since the measure of the boundary $\CM\back\CM_0$ is zero with
respect to any borelian measure one might be tempted to 
consider that from quantum viewpoint the effects of the fields which
do not belong to $\CA_0$ are negligible. From a functional integral
point of view their contribution is certainly negligible but in the
Hamiltonian formalism the fact that physical states must satisfy
some  boundary conditions just precisely at those field configurations
has a deep significance for the low energy behaviour of the theory.

\newsec{Maximal Non-abelian Holomorphic Gauge}

One consequence of the previous analysis is that, for the gauge orbits
of two-dimensional gauge fields in $\CM_0$ which are associated to trivial
holomorphic bundles,  it is possible to find a global gauge
fixing condition which is free of Gribov ambiguities. The basic idea
is that these fields can be globally written as $A_{\bar z}=h^{-1}
\partial_{\bar z} h$ in terms of a complex
scalar field $h$ with values on SL(N,$\IC$) and any matrix of SL(N,$\IC$)
can be split by the polar decomposition  $h=HU$ as the product of a
positive hermitean matrix $H\in  \Cm^+(N,\IC)$  and a unitary matrix $U\in
SU(N)$, both with unit determinant.  $H$  can be
identified with the positive square root of the positive operator $h
h^{\dag}$ (i.e.  $H=\sqrt{h h^{\dag}}$)  and $U=H^{-1}h$.

Using such a decomposition any gauge field $A\in \CA_0$ can be
rewritten as 
\eqn\der{A_{\bar z}=U^{\dag} ( H^{-1}\partial_{\bar z} 
H) U+ U^{\dag}\partial_{\bar z} U}
 which means that $A$ is gauge
equivalent to the gauge field   
\eqn\fin{A^U_{\bar z}=
H^{-1}\partial_{\bar z} H,}
because \der\ implies that 
$$A^U = U A U^{\dag} + U d U^{\dag}.$$ 
Therefore, the orbit space  of  gauge
fields  which are associated to trivial holomorphic bundles, $\CM_0^{S^2}$, 
is in one-to-one correspondence with the space of
pointed functionals from $S^2$ into the space of positive definite
hermitean matrices with unit determinant, $H:S^2 \to \Cm^+(N,\IC)$,
i.e.  
$$\CM_0^{S^2} \equiv \Maps(S^2, \Cm^+(N,\IC)),$$
where the subscript indicates that the maps are trivial
at the north pole $x_n$ of $S^2$, $H(x_n)=\II$.

The Maximal Non-Abelian Gauge  is
defined by the condition
\eqn\mnagc{A_{\bar z}=
H^{-1}\partial_{\bar z} H.}
Notice that all gauge fields of the form \mnagc\ are gauge
inequivalent, thus the  Maximal Non-Abelian Gauge is free of Gribov
ambiguities. The gauge orbits of these fields under the  group of
pointed gauge transformations $\CG_0$ fill the open dense 
subset $\CA_0$ of the whole space $\CA$  of gauge fields over $S^2$ and the
parametrization \mnagc\ is a complete gauge fixing condition for
them.  This implies that the domain of gauge orbits $\CM_0^{S^2}$
parametrized by the trivial holomorphic gauge fields is dense in the
whole orbit space $\CM_{S^2}$. In this sense the Maximal Holomorphic
Gauge condition is complete. This also implies that $\CM_0^{S^2}$ is
contractible because the space of positive hermitean matrices  with
unit determinant is  open and contractible. The cohomology groups of
$\CM_0^{S^2}$ are, thus, trivial $H^n(\CM_0^{S^2},\IR)=0$ for $n>0$.
This can be  explicitly checked by showing that all the non-trivial
closed forms of $\CM_{S^2}$ become pure differentials or exact on
$\CM^{S^2}_0$. 
The cohomology and
homotopy of  $\CM_{S^2}$ are  encoded in the boundary
$\CM^{S^2}_\ast=\CM^{S^2}\back\CM_0^{S^2}$ of $\CM_0^{S^2}$.
Moreover, the orbits belonging to this boundary can easily be
identified because the gauge fields $\CA$ of those orbits satisfy the
following equivalent properties  \unst,

\item{i)} the Dirac operator $\dsl_A$ has zero modes.  
\item{ii)}  for any map $h: S^2\back\{x_n\}\to SL(N,\IC)$ such that
$A_{\bar z}=h^{-1}\partial_{\bar z} h$ the
Wess-Zumino-Witten action $S_{\rm WZW}(h)$ 
diverges.  
\item{iii)} the holomorphic bundle
associated to $A$ is unstable  according to Mumford's definition
\item{iv)} the operator $\Dsl_{\bar z}$ has pointed zero modes on the
adjoint bundle, i.e. there are global holomorphic  sections of this
bundle  vanishing at one point of the sphere.

Moreover gauge fields of the boundary of
$\CM_0$ can be explicitly identified with well known
gauge fields. Let us
consider the abelian magnetic monopole in $SU(2)$ theory, 
\eqn\mm{A^{\rm mon}_{\bar z}=\varphi_-^{-{1}}\partial_{\bar z}
\varphi_-=  \varphi_+^{-{1}}\partial_{\bar z} \varphi_+= {1\over
(1+|z|^2)^2}\pmatrix {- z & -1\cr z^2 & z},} 
with
$$\varphi_-=\pmatrix {z & 1\cr {-  (1+|z|^2)^{-1}}&{\bar {z}
(1+|z|^2)^{-1}}};\, \varphi_+=\pmatrix{ 1 & {1/ z} \cr {-z
(1+|z|^2)^{-1}} &{|z|^2 (1+|z|^2)^{-1}}}.$$

It is obvious that   $A^{\rm mon}$ 
can never be globally  expressed as $h^{-1}\partial_{\bar z} h$. In
fact, it can be shown that the transition function connecting  two
holomorphic patches associated to the north and south hemispheres is
of the form 
$$g_{12}(x)= \pmatrix{z^{}&0\cr 0&1/z\cr},$$ 
which means
that the corresponding configurations can be associated with
monopoles of magnetic charges $\pm 1$. This shows that $[A^{\rm
mon}] \in \CM^{S^2}_\ast=\CM^{S^2}\back\CM^{S^2}_0$. The traceless
character of $\su\n$ means that properly speaking there is no net
magnetic charge for SU(N) gauge fields. However, for abelian SU(N)
gauge fields this vanishing condition can be satisfied in two
different ways: either A is made of elementary U(1) magnetic
monopoles with positive and negative magnetic charges which 
cancel each other out, or A does not
contain magnetic monopoles at all. The field \mm\ corresponds to the 
first type of fields. It is, thus,  not inappropriate to consider  these kind
of configurations  as non-abelian magnetic monopoles;
and as we shall see later this notion can be extended for arbitrary
non-abelian gauge fields.

There is another indication that the orbit of $A^{\rm mon}$ belongs 
to $\CM^{S^2}_\ast$.
It is the existence of pointed holomorphic sections on the 
adjoint bundle. 
From the  four independent holomorphic sections 
$\chi_{(\mu)};\, \mu=0,1,2, 3$ of the adjoint bundle  
$$
\eqalign{ \chi_{(3)}= & {1\over 1+|z|^2} \pmatrix{ |z|^2-1 &  2
{\bar z} \cr 2 z & 1-|z|^2 }\cr \chi_{(k)}= & {z^k\over (1+|z|^2)^2} 
\pmatrix{ \bar z & -{\bar z}^2 \crcr 1 & - \bar z}\qquad\quad
k=0,1,2\cr} $$ 
only two vanish at $z=\infty$ ($\chi_{(0)}$ and $\chi_{(1)}$). Then, the 
group of  pointed complex (chiral) gauge transformations which leave 
the holomorphic bundle associated to  $A^{\rm
mon}$ invariant is two-dimensional, i.e.
$$\dim\{h:S^2\longrightarrow SL(N,\IC), 
h(\infty)=\II; A_{\rm mon}^{h}=A_{\rm mon} \}=2.$$
This means
that the isotopy group of $\CG^\ic_0$ for this bundle is 
two-dimensional and that the codimension of the corresponding  
$\CG^\ic_0$--orbit has (real) codimension four in the space of all gauge
fields $\CA$. This is in contrast with the  orbit  generated
by the full group of complex gauge transformations which has 
codimension two. The difference is covered by the the non-trivial
$\CG^\ic_0$ moduli space of dimension two of unstable bundles of monopoles.
Notice that the bundles of this two dimensional moduli space are equivalent
with respect to the group of complex gauge transformations $\CG^\ic$.

Thus, the boundary of the domain $\CM_0^{S^2}$ where
the Maximal Non-Abelian Holomorphic gauge is well defined is the
closure $\overline\CM_1^{S^2}$ of the orbit space $\CM_1^{S^2}$ of
gauge fields which are holomorphically equivalent to the monopole \mm\
by   complex gauge transformations of $\CG^\ic$. This makes possible
 to extend the
gauge fixing condition to a larger domain 
$\CM_0^{S^2}\cup \CM_1^{S^2}$  in $\CM^{S^2}$. The orbits of
$\CM_1^{S^2}$ can be parametrized in terms of
maps $H:S^2 \to \Cm^+(N,\IC)$ by the gauge condition
\eqn\lar{A= H^{-1}A^{\rm mon}H +H^{-1} d  H.}
Notice that  $\chi_{(j)}\ j=0,1,2,3$   
generate  one-parametric subgroups of the gauge group of
complex gauge transformations which leave $A^{\rm mon}$ invariant.
Therefore the parametrization \lar\ is not unique unless we mod out
by those subgroups. This fact also explains why $\CM_1^{S^2}$ has
codimension 2 in $\CM^{S^2}$ whereas $\CM_0^{S^2}$ had codimension
0. The procedure could be extended in similar way for higher magnetic
monopole field configurations giving rise to parametrizations of
larger subsets of gauge fields orbits. However, the different charts
cover disjoint subsets of $\CM^{S^2}$ with different codimensions,
and they can not be considered as a single gauge condition.

The above construction of the Maximal Non-Abelian Holomorphic
gauge condition can be generalized for higher dimensional spaces.
For simplicity,  we only consider  gauge fields  $A$
defined on a trivial bundle $S^d\times SU(N)$, although the
extension of the  results for more general cases is straightforward.
In such a case a gauge field $A$ can be identified with a 
$\su\n$--valued one form over $S^d$.  

	Let us first introduce a very special coordinate system on $S^d$.
If $x_j,j=1,\cdots, d$ are the cartesian coordinates of
$\IR^{d}$, we define the angular coordinates 
$\varphi_j=2\arctan (x_j/2), j=1,\cdots, d$ which compactify
$\IR^{d}$ into a torus $T^{d}$. Now, if we exclude the
north pole $x_n$ of the sphere it can be
identified with $\IR^{d}$   by means of the stereographic
projection  $\pi_s$ from the north pole $x_n\in S^d$. Thus, the
coordinates $\{\varphi_j\in (-\pi,\pi); j=1,\cdots, d\}$ define a
complete set of orthogonal coordinates on $S^d\back\{x_n\}$. The
pullback by the stereographic projection $\pi_s$ of the hyperplane
$$
\pi^{-1}\{
(\varphi_1=\varphi,\varphi_2,\cdots,\varphi_d);  \varphi_j\in [
-\pi,\pi); j=2,\cdots, d\}
$$
of $\IR^d$ defined by the condition 
$\varphi_1=\varphi\in [-\pi, \pi)$ generates a
$S_\varphi^{d-1}$--sphere in $S^d$ which reduces at a single point
$x_n$ for $\varphi=\pm\pi$.

\vskip 0.2cm
 {\hskip1,4cm \epsfxsize=10cm \epsfbox{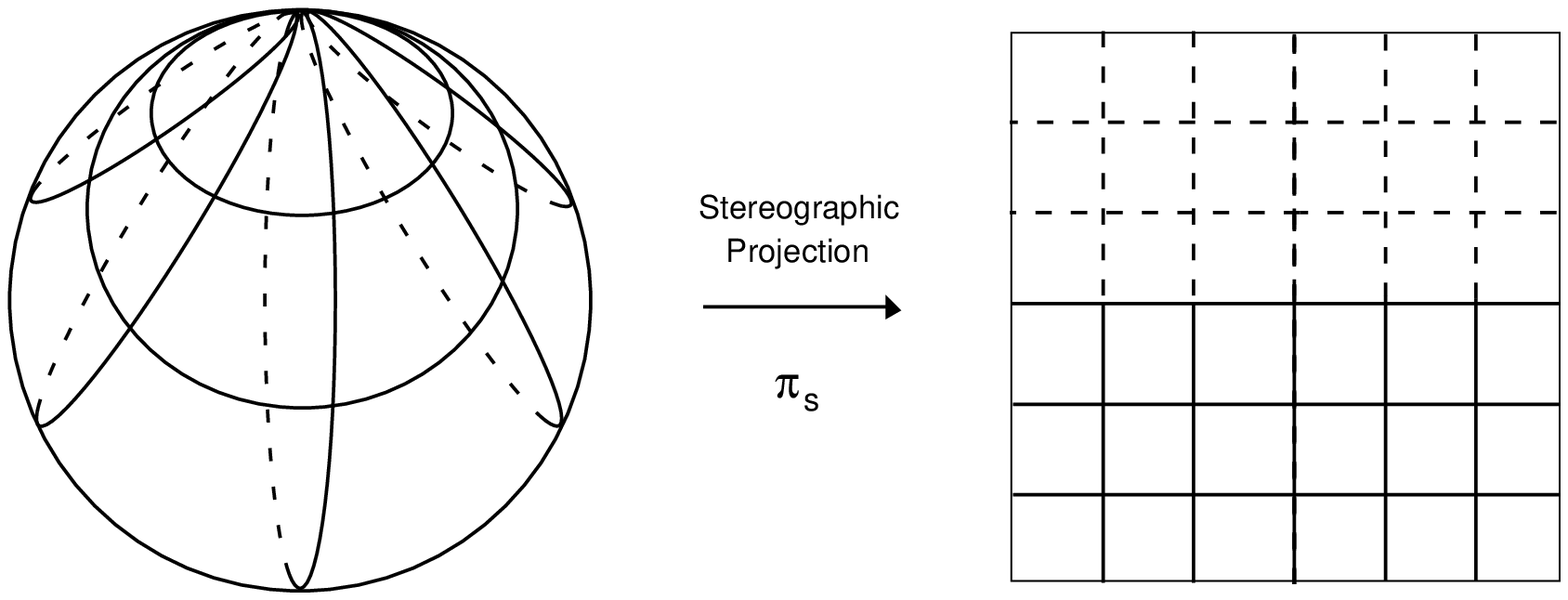}}
\vskip 0.1cm
\centerline{\hbox{\sevenpoint{ \bf Figure 1.} }}
\vskip -0.1cm
\centerline{\hbox{\sevenpoint 
Coordinates of the $S^d$--sphere induced by  stereographic projection
from the cartesian coordinates of $R^d$ }}
\vskip 0.3cm
 \tenpoint

Finally,   we normalize
 the radius $R_\varphi=\cos\varphi/2$ of the different
spheres  $S^{d-1}_\varphi$--spheres  to unit which means that the
corresponding embedding  $\j_{\varphi}$ of $S^{d-1}$  into $S^d$
becomes singular at $\varphi=\pm \pi$. The pullback by
$\j_{\varphi}$ of any gauge field $A$ on $S^d$ defines a loop of
gauge fields on $S^{d-1}$, 
 \eqn\dr{A(\varphi)=\j_{\varphi}^\ast A}  
with the same gauge group SU(N). In the extreme cases $\varphi=\pm
\pi$ the induced gauge field becomes trivial $A(\pm \pi)=0$.

On the same way, $A\in\CA^{S^d}$ defines a loop of Higgs fields
\eqn\drr{\Phi(\varphi)= A_1(\varphi)=A\left(\partial_{\varphi_1}\right)}  
over $S^{d-1}$ with values on  $\su\n$, the
Lie algebra of SU(N). Let us denote by
\eqn\higgss{\CH^{S^{n}}=\{ \phi:S^n\to {\rm ad\,} P(S^n,SU(N))\}} 
the space of Higgs fields defined as sections of the adjoint bundle
of a SU(N)  principal bundle $P(S^n,SU(N))$ over the
$S^n$--sphere.  In our case, these bundles
$P_\varphi(S^{d-1},SU(N))$ are trivial because the original bundle
$P(S^{d},SU(N))$ was assumed to be trivial. Then, $\CH^{S^{n}}=\{
\phi:S^n\longrightarrow \su\n$.
The above construction shows that there is  map 
\eqn\loopa{\CA^{S^d}\longrightarrow
\Maps(S^1,\CH^{S^{d-1}}\times \CA^{S^{d-1}})} 
defined by  the loops of  fields 
$$A\mapsto \left(A(\varphi),
\Phi(\varphi)\right) \in \CA^{S^{d-1}}\times
 \CH^{S^{d-1}}\qquad \varphi\in [-\pi,\pi] $$ which satisfy the boundary
conditions
\eqn\bcon{A(\pm\pi)=0 \qquad \Phi(\pm\pi)=\Phi_0
\prod_{j=2}^{d} \cos \varphi_j/2.} 
In
fact, because the map \loopa\ defines  a one-to-one
correspondence it is possible to reconstruct from the data
$\left(A(\varphi), \Phi(\varphi)\right)\in\CA^{S^{d-1}}\times
\CH^{S^{d-1}}$  the  original
$S^d$--gauge field $A$. Modding out by the group
of gauge transformations we obtain a one to one  correspondence   
\eqn\loopab{\CM^{S^d}\longleftrightarrow
\Maps(S^1,\CH^{S^{d-1}}\times \CM^{S^{d-1}})} 
between the
 orbit space $\CM^{S^d}$ and the pointed  loop space of 
$\CH^{S^{d-1}}\times \CM^{S^{d-1}}.$ Iterating this procedure we can
establish the following sequence of one-to-one correspondences 
\eqn\loopabc{\CM^{S^d}\longleftrightarrow \Maps(S^n,\CH^{S^{d-n}}
\times \CH^{S^{d-n}}\times  \bupnn \, \times \CH^{S^{d-n}}\times
\CM^{S^{d-n}})} 
for any positive integer $n\leq d-1$.  Obviously,
these  new characterizations of the space of gauge orbits preserve
the non-trivial topological structure of $\CM^{S^d}$. In particular,
it is straightforward to check that homotopy and cohomology groups are
identical to those of $\CM^{S^d}$ displayed in Section 2.

One of those characterizations offers an special interest for the
construction of the maximal gauge. It is the characterization based
in the 2--dimensional gauge fields orbit space $\CM^{S^2}$, i.e. 
\eqn\loopabcd{\CM^{S^d}\longleftrightarrow \Maps(S^{d-2},\CH^{S^{2}}
\times \CH^{S^{2}}\times \buppd \,\times \CH^{S^{2}}\times
\CM^{S^{2}}),} 
In particular, for $d=3$ one extra Higgs field is
enough to describe the gauge orbit space in terms of 2--dimensional
gauge fields. In four dimensions the same construction requires two
Higgs fields. Notice, that since the Higgs sector is topologically
trivial and all  interesting topological properties are encoded in
this picture by the structure of the space $\Maps(S^{d-2},
\CM^{S^{2}})$. 

For non-trivial bundles the generalization is
obvious but then the boundary conditions \bcon\ change in a way that
the correspondences do not lead to pointed maps from $S^n$  into the
space of gauge fields over lower dimensional spaces \loopa\ 
but the description  in terms pointed closed maps from 
$S^n$ into the  orbits spaces \loopabc\
remains  a one-to-one correspondence.

Now, we know from  Section 3 the structure of 
$\CM^{S^{2}},$then, we can define in analogy with what was done
there a Maximal Non-Abelian Holomorphic gauge for the gauge fields
$[A]\in \CM^{S^{d}}_0$ which corresponds to maps of \loopabc\
which entirely lie on $\CM^{S^{2}}_0$, i.e.
$$\CM^{S^d}_0=\Maps(S^{d-2},\CH^{S^{2}} \times \CH^{S^{2}}\times 
\buppd \,\times \CH^{S^{2}}\times \CM_0^{S^{2}}).$$
 The gauge fields orbits 
in such a submanifold are characterized by means of the
correspondence  
\eqn\pf{A\mapsto
\left(\Phi_1,\Phi_2, \cdots, \Phi_{d-2},  H^{-1}\partial_{\bar z
}H\right)}
in terms of a set of
d-1 pointed functionals  $$(\Phi_1,\cdots, \Phi_{d-2};H)\in
[\Maps(S^{d-2},\CH^{S^2})]^{d-2}\times \Maps(S^2\times S^{d-2},
\Cm^+(N,\IC)) $$ of $S^2\times S^{d-2}$ with values
in $\su\n$ and $\Cm^+(N,\IC)$, respectively. 

This defines the Maximal Non-Abelian Holomorphic gauge condition
for any dimension.
 It is obvious that  fields satisfying the gauge condition \pf\ do
not have Gribov copies under the group of pointed gauge
transformations $\CG_0=\{U\in\CG;U(x_n)=\II\}$.
From the definition it follows that $\CM_0^{S^d}$ can be
essentially parametrized in terms of a system of affine coordinates in
$\Maps(S^2,\Cm^+(N,\IC))$ and $[\CH^{S^2}]^{d-2}$.
This explicitely shows that the fundamental domain  $\CM^{S^d}_0$ 
of the Maximal Non-Abelian Holomorphic gauge is a contractible
submanifold of the space of gauge orbits $\CM^{S^d}$. The 
topologically non-trivial sector of $\CM^{S^d}$ is  essentially
equivalent to $\Maps(S^2,\Cm^+(N,\IC) $ because of the triviality of 
the Higgs fields sector.

Now, the efficiency of the gauge fixing condition decreases as
space--time dimension increases. In  three dimensions 
$\CM^{S^3}_0$ constitute an open  dense set on $\CM^{S^3}$, because
$\CM^{S^2}\back\CM^{S^2}_0$ is a codimension two manifold in
$\CM^{S^2}$. Thus, the space of loops which do not reach the
manifold  $\CM^{S^2}\back\CM^{S^2}_0$ is open and dense in the space
of all pointed loops which means by the correspondence 
\loopabcd\ that $\CM^{S^3}_0$ is an open and dense 
submanifold of the whole space of orbits $\CM^{S^3}$.
 In four dimensions,
however, the space $\Maps(S^2,\CM^{S^2}_0)$ is not
dense in  $\Maps(S^2, \CM^{S^2})$, because  if a map reaches the
dense submanifold  $\CM^{S^2}\back\CM^{S^2}_0$,
 generically it cannot be transformed by an infinitesimal
transformation into another one that does not intersect such a
submanifold. In particular this implies  that the gauge fixing \pf\
is not complete because there is an open  set of gauge orbits which
do not intersect the gauge condition slice. Moreover, if we consider
  a non-trivial bundle $P(S^d,SU(N))$ none of the gauge fields with
non-trivial topological charge induce a map in
$\Maps(S^2,\CM^{S^2})$ which does not intersects 
$\CM^{S^2}\back\CM^{S^2}_0$. The situation
gets even worse for dimensions $d>4$.

However, the fact that gauge condition \pf\ is complete for
$d=3$ is very important because the Yang-Mills theory in 4+1
dimensions can be described in the Hamiltonian formalism in terms of
3-dimensional gauge fields. Moreover, as will be analysed in 
section 7, the fact that there are many four-dimensional gauge configurations
which can not be described by the maximal gauge condition might have
relevant implications for the  understanding of confinement in the
dual superconductor picture 
\ref\gth{ 't
Hooft, In {\it Gauge Theories with Unified Weak Electromagnetic and
Strong Interactions},  High Energy Physics, eds. A. Zichichi,
Compositori, Bologna (1976)\semi Nucl. Phys.
{\bf B190}  (1981) 455}\ref\mand{ S. Mandelstam, Phys. Rep.
{\bf 23}  (1976) 245; Phys. Rep. 
 {\bf 67} (1980) 109}.

On the other hand, the characterization of gauge fields orbits in a d
dimensional space-time in terms of pointed maps from an $S^{d-2}$ sphere
into  $[\CH^{S^{2}}]^{d-2}\times \CM^{2}$ is always one-to-one. The problem is
that one would like to have a parametrization by affine
coordinates of a maximal open  subset in such a space. With the
gauge fixing condition described above we only can achieve a complete
parametrization on the cases $d\leq 3$. It is not, however, excluded
the existence of  another gauge in higher dimensions covering a larger
subset of gauge orbit spaces. In fact, this is possible if we exclude
the requirement of having a uniform parametrization by affine
coordinates of the same dimensionality,  which seems to be necessary
to have a correct  particle interpretation of the physical spectrum.
Indeed, to describe fields beyond the horizon of $\CM^{S^d}$ 
 we have to consider  maps in $\Maps(S^{d-2},\CM^{S^2})$ which 
reach  the stratum $\CM^{S^2}_1$. But, since the orbits of 
$\CM^{S^2}_1$  can also be parametrized  by \lar\  
a larger set of gauge field orbits over $S^d$ can be parametrized
along the same lines.
Since the codimension of $\CM^{S^2}\back (\CM^{S^2}_0\cup
\CM^{S^2}_1)$ is six the set  $\Maps(S^{d-2},\CM^{S^2}_0\cup
\CM^{S^2}_1)$ is not only an open subset of $\Maps(S^{d-2},\CM^{S^2})$
but its boundary has codimension 8-d which implies that it is also
dense for $d<8$. The only problem is that the parametrization of the
corresponding open dense domain in the orbit space  $\CM^{S^d}$ is achieved
in terms of two sets of affine coordinates with different
dimensions.

To some extend this analysis shows from a different perspective
the reasons why it is impossible to cover the whole space of orbits
in the non-abelian case. 
In principle, the problem of the existence of
obstructions to a complete gauge fixing looks like a  technical
problem, but
the observed difference  between the $d\leq 3$  and  $d> 3$ cases
might have a physical meaningful interpretation if as  suggested below  the
confinement mechanism  is related to this kind of topological
obstruction. Notice that in $d\leq3$, perturbation theory always leads
to a confining potential   which only monopoles in $d=3$ improve a
little bit  from being logarithmic to become linear. 
We will see that those  configurations  belong to the
boundary $\CM^{S^3}_\ast$ of  $\CM^{S^3}_0$ which is the first
indication that the configurations with lie beyond the boundary of
$\CM^{S^d}_0$ might play an special role in the confinement mechanism. In
  Section  7 we will see that these configurations do carry a
kind of non-abelian magnetic charge and therefore might represent the
intrinsic realization of the non-abelian monopoles which will drive
the vacuum structure to that one suited for dual superconductor
picture.

\newsec{One Dimensional Gauge Fields}

One could proceed further in the dimensional reduction mechanism to
get another characterization of a  contractible dense submanifold in
the space of gauge orbits.
 
Iterating once more the above  procedure we can establish one more
one-to-one correspondence 
\eqn\loopabcde{\CM_{S^d}\longleftrightarrow
\Maps(S^{d-1},\CH^{S^{1}} \times \CH^{S^{1}}\times \buppu \,\times
\CH^{S^{1}}\times \CM^{S^{1}}).} 
The non-trivial topological
sector is encoded in $\Maps(S^{d-1},\CM^{S^{1}})$ which can be easily
analysed because the pointed orbit space over the circle  is
isomorphic to the gauge group SU(N) itself. It is easy to check that
the topology of $\CM$ does coincide with that of 
$\Maps(S^{d-1},\CM^{S^{1}})$. In that  parametrization is clear what
the real obstruction to a global gauge fixing is: the topological 
structure of the gauge group SU(N). The problem of  finding a maximal gauge
of the $S^d$--dimensional gauge theory is reduced to 
find  a maximal gauge
with affine coordinates   
in a contractible  open dense subset $\CM^{S^{1}}_0$ of 
the orbit space of one-dimensional
gauge fields $\CM^{S^{1}}$ following the same steps  as in the case of 
maximal holomorphic gauge. It is then necessary to select
a domain in the one dimensional gauge orbit space 
$\CM^{S^{1}}\equiv SU(N)$ with a convenient parametrization 
by affine coordinates. This is equivalent to find an open dense 
set in SU(N) with those properties.

One subset of
unitary matrices which is contractible, open and dense in 
$\CM^{S^{1}}=SU(N)$  is
\eqn\thrid{\CM^{S^{1}}_0=SU(N)_0 =\{U\in SU(N); {\rm If}\
\dim\ker(U-{\rm e}^{i\alpha}\II)>1,
|\alpha| < {2\pi\over N}\}.} 
Its boundary
\eqn\thrid{\CM^{S^{1}}_\ast=
\CM^{S^{1}}\back \CM^{S^{1}}_0=SU(N)_\ast =\{U\in SU(N); {\rm If}\
\dim\ker(U-{\rm e}^{i\alpha}\II)>1, |\alpha|
\geq {2\pi\over N}\},} 
is a stratified space whose larger strata, given by the matrices 
of SU(N)$_\ast$
 with only double degenerated spectrum, has codimension three.
  The natural system  of affine coordinates  
for $\CM^{S^{1}}_0$ is given by the exponential  map in SU(N).
The Lie algebra $\su\n$ of   SU(N)   
defined by  traceless $N\times N$ hermitean matrices,
is completely covered by the
coadjoint orbits of the Cartan subalgebra of traceless diagonal real 
matrices $\hn$ of $su\n$ 

$$D=\pmatrix{\alpha_{1}&\cdots &0&0&\cdots & 0\cr
\cdot &\cdots &\cdot &\cdot &\cdots &\cdot\cr
0&\cdots &\alpha_{k}&0&\cdots &0\cr 
0&\cdots&0&\beta_{1}&\cdots & 0\cr
\cdot &\cdots &\cdot &\cdot &\cdots &\cdot\cr
0&0&\cdots&0&\cdots &\beta_{N-k}\cr};\quad \tr D=0.$$
Let us consider the open dense subset $\h_0$
in the Cartan subalgebra $\hn$ of $\su\n$  defined by
\eqn\esp{\eqalign{\h_0=&\left\{D\in \hn; {\rm with }\
\alpha_i\in[0,2\pi), \beta_j\in (-2 \pi,0); 
\sum_{i=1}^k \alpha_i<2\pi, \right. \cr & 2\pi+\sum_{i=1}^k 
\alpha_i +\beta_1\geq 2\pi;
 \alpha_1\leq\cdots\leq
\alpha_k\leq 2\pi+\beta_1\leq\cdots\leq2\pi+\beta_{N-k};\cr   
 & \left. {\rm and}\  \alpha_i<{2\pi\over N}\  {\rm if} \
\alpha_i=\alpha_j, {\rm and}\ \beta_i>-{2\pi\over N}
\ {\rm if} \ \beta_i=\beta_j \right\}.}}
The  subset $\su_0\n$  defined by the
coadjoint orbits of
$\h_0$,
 $$\su_0\n=\{V\in \su\n; V=U^{\dag} D U, {\rm for}\ D\in \h_0 ,U\in
SU(N)\},$$ 
is a contractible open  dense subset of $\su\n$.
It is trivial to see that the
restriction of  the exponential map to $\su_0\n$ establishes a  
one-to-one correspondence between  $\su_0\n$ and   
$\CM^{S^{1}}_0$.
This method  provides a unique and unambiguous parametrization of
the submanifold $\CM^{S^{1}}_0$  via the
exponential map and shows that $\CM^{S^{1}}_0$ is a
maximal contractible open subset of the space of gauge orbits of
one--dimensional gauge fields $\CM^{S^{1}}$.  
Coordinates of $\CM^{S^{1}}_0$ can be defined by
 the parameters of diagonal matrices in $\h_0$ and the  angular
variables of the coset $SU(N)/T_D$  where the unitary matrices $T_D$
leaving invariant the matrix $D$ have been modded out.
The coordinates are not well
defined for matrices of $\CM^{S^{1}}_0$ with  double degeneracy
which have a larger isotopy group $T_D$, e.g for the identity
$D=\II\in \su\n$  the isotopy group $T_D$ is the
full group SU(N). This {\it angular} coordinate system becomes
singular for those particular matrices 
in a similar manner as  the origin is singular for  polar
coordinates of the plane.  However, it is possible to choose
another set of affine coordinates which is non-singular in the 
subset $\su_0\n$ and  provides a  complete unambiguous affine
coordinate system for $\CM^{S^{1}}_0$. The inclusion of gauge fields
with degenerated eigenvalues in the fundamental 
 domain $\CM^{S^{1}}_0$ is necessary to
describe the classical vacua $A_{\rm vac}=0$ and define a maximal
open dense set around it. The geometric discussion  would be simpler
if we exclude those configurations and restrict the whole construction
 to the exponential of the Weyl alcove of
 the Lie-algebra. The
singularities associated to double degeneracies can be considered
then as defects and a proper counting of the associated magnetic
charges provides a description of the topological charge in terms of
those magnetic charges \ref\cj{N. Christ and R. Jackiw, Phys. Lett. 
{\bf B 91 } (1980) 228}--\nref\hr{H. Reinhardt, Nucl. Phys. {\bf B 503}
(1997) 505} \ref\wipf{C. Ford, T. Tok and A. Wipf, hep-th/9809209}, 
although it is not very satisfactory from a physically point of view  
to associate a magnetic charge to points where the gauge field vanishes. 
Our approach solves all
those potential problems by the choice of the gauge orbit space of
the  group of pointed gauge transformations. In the particular
case of $d=1$ the gauge condition becomes equivalent to Landau gauge. Thus,
the above construction explicitly shows the validity of the generalization of
the Semenov-Tyan-Shanskii-Franke result for $C^\infty$-smooth gauge
fields.

For U(N) gauge fields  there exists a similar maximal
open subset $\CM^{S^{1}}_0$. It is defined by $\CM^{S^{1}}_0=\{U\in
U(N); \det (U+I)=0\}$ and it is parametrized via exponential map by
the set of hermitean matrices with eigenvalues  $\alpha\in
(-\pi,\pi)$. The boundary of $\CM^{S^{1}}_0$ has generically
codimension one in the space of gauge orbits of one--dimensional
U(N) gauge fields. This parametrization explicitly  shows the nature of the
topological obstruction to get a complete gauge fixing condition
because in this case we have $\pi_1(\CM^{S^{1}})=\pi_1(U(N))=\IZ$.

\newsec{Maximal Non-abelian  $\sigma$--gauge}

Let us return to SU(N) gauge theories.
Since  $\CM^{S^{1}}_\ast$ has generically codimension three the
subset $\CM^{S^{d}}_0$ of $\CM^{S^{d}}$, defined by the
the gauge field orbits whose associated maps  
\eqn\loopabcdef{\CM^{S^d}_0= \Maps(S^{d-1},\CH^{S^{1}} \times
\CH^{S^{1}}\times \buppu \,\times \CH^{S^{1}}\times \CM^{S^{1}}_0),}
in the correspondence 
\loopabcde\ do not intersect the set $\CM^{S^{1}}_\ast$,
is open and dense in the space of all gauge orbits 
only for dimensions $d\leq 3$. This is
in agreement with the behaviour pointed out in the  Section 4
for the maximal holomorphic gauge. In this sense the obstructions to
the extension of both gauge conditions are compatible. However, they
are not completely  identical as the following example points out.
        Let us consider the abelian U(1) gauge theory over a two
dimensional sphere $S^2$. In that case we have 
\eqn\abel{\CM^{S^2}\longleftrightarrow \Maps(S^{1},\CH^{S^{1}}
\times  \CM^{S^{1}}),} 
which contains  an open contractible subset
given by the maps 
\eqn\abelb{\CM^{S^2}_0\equiv
\Maps(S^{1},\CH^{S^{1}}_0 \times  \CM^{S^{1}}_0),} 
where 
$\CM^{S^{1}}_0 = \CM^{S^{1}}\back \CM^{S^{1}}_\ast$ and $
\CM^{S^{1}}_\ast=  \{-\II\}$.  In this way we get a gauge
fixing for gauge fields associated to the maps which do not
intersect $\CM^{S^{1}}_\ast$, but this subset is not
 very large. For instance, it we consider U(1) gauge fields over
$S^2$  with non--trivial magnetic charge ($c_1(A)\neq 0$) every
associated map intersects $\CM^{S^{1}}_0$. Moreover, in the zero
magnetic charge sector there is a open set of maps which intersect
$\CM^{S^{1}}_0$ because $\CM^{S^{1}}_\ast$ has codimension one in
$\CM^{S^{1}}\equiv U(1)$. Of course the maps
described by the gauge condition have zero winding number but not
all maps with winding number zero are in $\CM^{S^{2}}_0$. This
restriction occurs in spite of the fact that the whole set of maps
with zero winding number is a
contractible manifold. In fact, this set is topologically equivalent
to the two-dimensional orbit space for trivial bundles 
$\CM^{S^{1}}$ which is also a contractible manifold. It is this
contractibility property what makes possible the application of the
method described in the  Section 4  and provides  a complete
gauge fixing condition for $\CM^{S^{2}}_0({\rm holomorphic})$. 
Any gauge field orbit in
that space has one and only one representative of the form 
$$
A_{\bar z}= H^{-1}\partial_{\bar z} H
$$
where $H:S^2\to \IR^+$ is any
positive real function over $S^2$. In this sense the
maximal holomorphic gauge goes beyond the maximal $\sigma$--gauge,
i.e. $\CM^{S^{2}}_0(\sigma)\subset \CM^{S^{2}}_0({\rm hol})\subseteq
\CM^{S^{2}}$. 

One could think that something similar might occur for four
dimensional gauge fields. Finding a better gauge in three dimensions
defined over a subset of gauge orbits larger than  $\CM^{S^3}_0$
will allow to go beyond the non-dense subset $\CM^{S^4}_0$ of
$\CM^{S^4}$. This requires
 the introduction of a gauge condition in three dimensions for a dense domain
of gauge orbits with boundary of codimension higher than one.
However, the fact that the topology of $\Maps(S^3,SU(N))$ is
non-trivial even if they are restricted to maps with zero winding
number indicates that such a possibility does not occur as in the
abelian case, where the corresponding set was contractible. This feature
does not exclude, however, the possibility of
existence of a more efficient gauge
condition in four-dimensions, which can be achieved following the
lines indicated in Section 4 by adding  to $\CM^{S^{1}}_0$
the strata associated to double degenerated diagonal traceless
matrices.

It has been shown in Ref.  \ref\asomitt{M. Asorey and   P.K. Mitter,
 Ann. Inst. Poincar\'e {\bf 45} (1986) 61}  that the maps in
$\Maps(S^{d-1},\CH^{S^{1}} \times \CH^{S^{1}}\times \buppu \,\times
\CH^{S^{1}}\times \CM^{S^{1}})$  associated to connections defined
in non-trivial SU(N)  bundles are generically non-contractible. In
fact, in four dimensions  ($d=4$) the instanton number $c_2(A)$
equals the winding number of the corresponding map from $S^3$ into
$\CM^{S^{1}}= SU(N)$\asomitt.  For symmetric gauge fields the
corresponding maps may become degenerated but in any case they
always intersect the submanifold $\CM^{S^1}_\ast$\unst.  

It is remarkable the  analogy of the
 gauge condition analyzed in this section  and the temporal gauge.
It is well known that temporal gauges do not completely fix the
gauge because time independent gauge transformations still transform
the fields without leaving the temporal gauge fixing slice. This
potential pathology does not occur in our approach because all the
temporal Polyakov lines used to define the gauge condition
intersect at the point $x_n$ which excludes the existence of Gribov
copies under any kind of pointed gauge transformation. The remaining
gauge freedom under global gauge transformations only involves a
finite number of degrees of freedom which do not lead to any
infrared divergence because the weight of the redundant copies of
generic gauge fields is bounded by the  finite volume of the gauge
group SU(N). 

This characterization of SU(N) pure gauge fields in terms of maps
from $S^{d-1}$ into the gauge group SU(N) is reminiscent of
the low energy description of QCD in terms  of Chiral
models. This suggest that a gauge invariant description of
the physical degrees of freedom of pure Yang-Mills theory
can be achieved in similar terms. In that description glueballs
can be naturally identified in terms of the SU(N) sigma model
variables. This opens a new avenue to the description of the
low energy spectrum of pure gauge theories as an effective
theory described by an SU(N) sigma model.

\newsec{Beyond the Horizon: Non-abelian Monopoles and Confinement}

Although the main goal of this paper is to provide
a complete gauge fixing for a dense set $\CA_0$ of gauge fields
$\CA$ for $d< 4$, one might wonder whether 
the remaining gauge fields  $\CA\back\CA_0$ have  any physical relevance.
Indeed, they have zero measure  with respect to any
borelian functional  measure, but there are indications that they
can play a very relevant role in the non-perturbative dynamics of
gauge field theories.  In particular, for   physical effects which
depend on  topological  properties their contribution is of leading
order. For instance, they are responsible for the inconsistency of
pathological anomalous chiral theories. In the Hamiltonian
formalism, their role is enhanced because of the boundary conditions
that physical states must satisfy at $\CA\back\CA_0$, 
and we know that boundary conditions are very
important for the low energy behaviour of any quantum theory. On the
other hand, one might ask why a specific choice of
coordinates on the orbit space would be more relevant than another.
The answer comes from the fact that the Yang-Mills functional gives
a different weight to the classical configurations either in the
euclidean approach where it measures the contribution of the
different classical configurations or in the Hamiltonian where it
indicates which configurations are more relevant not only for the
classical dynamics but also for the quantum theory. In both cases
the leading configurations are the solutions of Yang-Mills equation,
irrespectively of their stable or unstable character. In the canonical 
approach the later provide a measure of quantum tunneling and
include sphalerons. In the covariant approach the unstable stationary
configurations mark the interphases of the different instanton liquids.  
We shall see that most of these configurations lie at the boundary of
the fundamental domain the Maximal Non-Abelian Gauge. 

Another interesting
problem which is related to those boundary configurations is
confinement. In the dual superconducting scenario for confinement
the basic ingredient is a monopole condensation. However,
it is commonly accepted that there is no intrinsic definition of a
magnetic monopole for SU(N) gauge fields. One of the appealing
characteristics of abelian projection is that it allows to identify
some configurations which carry in that representation a sort of
magnetic charge \th. Moreover, those configurations seem to play a key
role in confinement when they are considered in the maximal abelian
projection gauge. However, this concept of monopole is extremely
gauge dependent. 

We are now in a position of  providing an intrinsic
definition of such monopoles based in the above characterization of
gauge fields. We will say that a $S^d$ gauge field configuration
carries a non-abelian magnetic charge (i.e. it is made of monopoles)
when the corresponding map from $S^{d-2}$ into $\CM^{S^2}$ intersects
the orbits associated to unstable bundles
$\CM^{S^2}\back\CM^{S^2}_0$. The interpretation of configurations
of  $\CM^{S^2}\back\CM^{S^2}_0$ as non-abelian monopoles is quite
natural because the associated holomorphic bundles do admit abelian
subbundles with non-trivial first Chern classes. The
characterization is completely intrinsic it is only based on the
complex structure of $S^2$. For higher dimensional gauge fields the
generalization is obvious if we assume that to detect a magnetic
charge we need a sort of two-dimensional $S^2$ device  to measure
the magnetic flux leaving the enclosed domain which indicates the presence or
absence of  monopoles. This is what the above definition prescribes.

The only problematic aspect of this intrinsic concept
 of monopole is that it does not carry an extensive charge and
that the charge is not localized. In fact it is a quite subtle concept
of charge because for a given field configuration, a fixed
$S^2$ sphere inside $S^d$ can enclose a 2-d
monopole but a slight perturbation of it does not. However, this
evanescent aspect has also some advantages: it is possible to attain
gauge field configurations which describe a dense gas of monopoles
and they become generic in four-dimensions. They correspond just to 
the configurations which make our gauge fixing incomplete. The fact
that in 3-dimensions configurations without monopoles are generic
whereas in 4-dimensions they are not, suggests that those
configurations might play a relevant role in providing an effective
linear potential at large distances. Notice that in 3-dimensions the
change from the perturbative Coulombian logarithmic confining
potential to the real linear potential is not so substantial and in fact can
be achieved due to the mild contribution of point-like monopoles
which are at the boundary of $\CM_0$ \ref\poly{ A. M.  Polyakov,
Phys. Lett. 
 {\bf B 59}  (1975) 82; Nucl. Phys. 
{\bf 120}  (1977) 429}. A dilute gas of monopoles
picture is enough to describe the phenomenon. In four dimensions our
conjecture implies that the linear potential is likely to be
associated with the fluctuations of those monopole  configurations
which seem to have wilder field interactions promoting the role of
the configurations associated to a dense gas or liquid of
monopoles.

In order to understand the role  
boundary/monopole configurations in quantum gauge theories 
let us analyze some related
 quantum effects. 

In two-dimensional QCD  it is known that the fermionic determinant
vanishes for gauge field configurations of 
$\CM^{S^2}\back\CM^{S^2}_0$ \ref\mkp{ A.  Kupiainen and J. Mickelsson,
 Phys. Lett. {\bf B 185} (1987)  107-110}. In fact
this was  the first theory where it was realized the role of a
holomorphic gauge fixing. 

The physical relevance of those boundary configurations is not
exclusive of 1+1 dimensional gauge systems.
In 2+1 dimensional Yang-Mills theories on a finite volume, there
exist static solutions of Yang-Mills equations which are critical
points of the 2-dimensional Yang-Mills functional that is the
effective potential of the 2+1-dimensional theory. They are called
sphalerons and although they cannot   be stable they are unstable
in a minimal way. There is only   a finite number of instability 
decaying modes. On the other hand  the actual value of the
Yang-Mills functional on those saddle point configurations  marks
the height of the  potential barrier responsible of the existence of
relevant non-perturbative effects. These sphalerons were extensively
analyzed by Atiyah and Bott for any Riemann surface  and Lie
group  \ab. In our case  the Riemann surface   is the
two--dimensional  sphere $S^2$, and the Atiyah-Bott results show
that in such a case all sphalerons belong to the boundary  
$\CM^{S^2}\back\CM^{S^2}_0$. In fact for each class of unstable
bundles there exist only one different type of sphaleron solution. For
SU(2) the first non-trivial solution is the abelian magnetic
monopole \mm.

Those configurations  have leading role in the non-perturbative
dynamics in the Hamiltonian approach.  For instance, it is well
known that in the abelian case (QED$_{2+1}$)  when a compact
lattice regularization is introduced the logarithmic perturbative 
Coulomb  potential becomes linear by means of Debye screening of
electric charges in a monopole gas \poly\
 in a similar manner as vortices drive the
Berezinskii-Kosterlitz-Thouless phase transition in the XY model
\ref\bere{ V.L. Berezinskii,  Sov. Phys. JETP {\bf 32 } (1971) 493}
\ref\kt{ J.M. Kosterlitz and D.J.  Thouless, J. Phys.  {\bf C6}  (1973)
1181}. It is, therefore, natural to conjecture that the role of
configurations sitting at the boundary  $\CM^{S^2}\back\CM^{S^2}_0$ 
 which have been identified with the non-abelian generalization of
magnetic monopoles would play a similar role in the mechanism of
quark confinement of the non-abelian theory. This provides a
geometric setting for the 't Hooft-Mandelstam scenario \gth\mand\ in 2+1
dimensional gauge theories.

This conjecture is   supported by the fact that  those
monopole-like  boundary configurations become  extremely suppressed
in the vacuum state of  topologically massive gauge theories \am\
which is  a non-confining medium. In fact, in those theories the
vacuum functional exactly vanishes for such configurations  \ref\nod{M. Asorey,
F. Falceto, J.L. L\'opez and G. Luz\'on, 
 Phys. Lett. {\bf B 349} (1995) 125 }. 
The result follows from Ritz variational principle which
establishes that the expectation value  of the Hamiltonian on
physical states has to be minimized by the quantum vacuum. The
existence of vacuum nodes at gauge fields in the boundary of the
maximal  holomorphic gauge fixing condition is not only required for
the minimization of the kinetic term, but also for that of the
Yang-Mills potential term. Both terms, kinetic and potential, of the
Hamiltonian conspire to force the vanishing of the vacuum functional on gauge
fields which are on the complex  gauge orbits of the monopoles \mm,
i.e. the whole boundary   $\CM^{S^2}\back\CM^{S^2}_0$. Since 
for $k\neq 0$ the theory is not confining  it is foreseeable to associate
to the suppression of the fluctuations of those nodal configurations
a leading role in the breaking mechanism of quark confinement. Conversely,
 it is conceivable that those fluctuations could play a
relevant role in  the mechanism of quark confinement when $k=0$,
where vacuum nodes are not expected to appear according to Feynman's
qualitative arguments \ref\feyn{ R. Feynman, Nucl. Phys. 
 {\bf B 188}  (1981) 479}.

 In 3+1 dimensions we
also have relevant configurations carrying intrinsic monopole
charges. In a finite volume $S^3$--sphere there are sphaleron
solutions of Yang-Mills equations which measure the height of the
potential barrier between classical vacua and, therefore, the
transition temperature necessary for the appearance of direct
coalescence between those vacua. They also serve as indicators of
the relevance of non-perturbative contributions. For $SU(2)$  the
sphaleron in stereographic coordinates reads 
\vskip-.2cm  
\eqn\dos{
A^{\hbox{\sevenrm{sph}}}_j={4\rho\over (x^2+4 \rho^2)^2}(4\rho 
\epsilon^a _{jk} x^k + 2x^a
x_j- [x^2-4\rho^2]\delta^a_j)\sigma_a,}
where $\rho$ is the radius of the $S^3$ sphere. The unstable mode can
be identified with the deformation under scale transformations.

Now, the sphaleron \dos\ defined on the sphere of radius $\rho$ 
induces in the Maximal Non-Abelian Holomorphic gauge a loop of
2-dimensional gauge fields $\{\j^\ast_\varphi A,
-\pi<\varphi\leq\pi\}$ on the  $S^2$--sphere, which for $\varphi=0$ 
becomes gauge equivalent to the  
abelian Dirac monopole  gauge field \mm, i.e. 
$\j^\ast_0 A= \Phi^{-1} A^{\rm mon}\Phi$, with
$\Phi= {\rm exp}(-i\sigma_3 \pi/4)$.
This means that $[\j^\ast_0 A]\in \CM^{S^2}\back\CM^{S^2}_0$, and, therefore,
the sphaleron itself belongs to the boundary of the maximal domain of
the maximal non-abelian holomorphic gauge, i.e. 
$[A^{\hbox{\sevenrm{sph}}}]\in \CM^{S^3}\back\CM^{S^3}_0$. Once more
a relevant configuration for the non-perturbative behaviour of the
theory belongs to the boundary of the maximal domain of the gauge 
condition.

Another very relevant property of sphalerons are that they give a
very special value to the Chern-Simons functional 
\vskip-.2cm 
$$C_s(A^{\hbox{\sevenrm{sph}}})=
{1\over 4\pi}\int \tr\ \left(A^{\hbox{\sevenrm{sph}}}\wedge 
dA^{\hbox{\sevenrm{sph}}}+{2\over 3}A^{\hbox{\sevenrm{sph}}} 
\wedge A^{\hbox{\sevenrm{sph}}}\wedge
A^{\hbox{\sevenrm{sph}}}\right)=\pi.$$ 

\noindent  
This property together with the parity behaviour of sphalerons and
the fact that $[A^{\hbox{\sevenrm{sph}}}]$ belongs to the boundary of
$\CM^{S^3}_0$ implies that the vacuum state of Yang-Mills theory at
$\theta=\pi$ vanishes  for sphaleron gauge fields,  i.e.
$\psi_0(A_{\hbox{\sevenrm{sph}}})=0$  \ref\aff{ M. Asorey and F.
Falceto, Phys. Rev. Lett.  {\bf 77} (1996) 3074-3077}.
The same properties also imply that
parity symmetry is 
not  spontaneous broken and the 
vacuum state $\psi_0$ is parity even \aff.    The absence of
spontaneous breaking of parity for $\theta=0$ 
 \ref\vw{ C. Vafa and E. Witten, Nucl. Phys.
{\bf B 234} (1984) 173; Phys. Rev. Lett. {\bf 53}  (1984) 535} and
$\theta=\pi$ is based on different physical arguments, but in both
cases the configurations of the boundary of $\CM^{S^3}_0$ play a
leading role. 

The existence of nodes in
the theory at $\theta=\pi$ is in contrast with what happens at 
$\theta=0$ where there are not such nodes \feyn.
Since the theory is expected to deconfine for $\theta=\pi$, the
result suggest that those nodes might be again responsible for the
confining properties of the vacuum in absence of $\theta$ term where
the vacuum has no classical nodal configurations.

In 3+1 dimensions we also have instantons which are the
 main responsible of the
non-perturbative contributions associated to tunnel effects between
classical vacua. Their effect seems to be very similar to that of monopoles in
compact QED in three-dimensional space-times. However, their contribution to
confinement does not seem to be crucial. Indeed, a standard argument
due to Witten shows that their contribution is exponentially
suppressed in the large N limit, whereas quark confinement is
strengthened in that limit  \ref\witt{ E. Witten,
 Nucl. Phys. {\bf  B149} (1979) 285}.  However, the instanton
contribution is very relevant for  the problem of chiral symmetry
breaking in the presence of dynamical  quarks \ref\dia{ D. Diakonov and
V. Petrov,
 Phys. Lett. {\bf 147B } (1984) 351; Sov. Phys. JETP {\bf 62 }(1985)
204, 431;   Nucl. Phys. {\bf B272 } (1986)  457}.
 Instantons with
unit topological charge   and structure group $SU(2)$  are given by 
\eqn\seisuno{A_\mu={2 \tau_{\mu \nu} (x-x_0)^\nu\over
(x-x_0)^2+\rho^2}} 
in stereographic coordinates of the four
dimensional  sphere $S^4$. 
There are two collective coordinates
which parametrize the moduli space of $k=1$ $SU(2)$ instantons: the
radius $\rho$ and its center $x_0$. The ${\goth sl}(2,\IC)$ 
matrices $\tau_{\mu\nu}= i( \tau_\nu^{\dagger} \tau_\mu -
\delta_{\mu\nu})/2$, with  $\tau=(-I,i\vec{\sigma})$ define a coupling
between  internal and external degrees of freedom.

In the maximal holomorphic picture they define a map $[\j^\ast A]$
from  $S^{2}$ into $\CM^{S^{2}}$. It can be shown that $[\j^\ast A]$ also
reaches the boundary of the maximal domain
$\CM^{S^2}\back\CM^{S^2}_0$ because the two-dimensional
gauge field $\j^\ast_{(0,0)} A$ 
\eqn\zero{(\j^\ast_{(0,0)} {A})_i= 
{1\over 2}{\epsilon_{ij}x^j \sigma_3\over
1+|x|^2},}
is gauge equivalent the abelian Dirac monopole
\mm\ with unit magnetic charge in $\CM^{S^2}\back\CM^{S^2}_0$.
Therefore, the instanton configuration also induces  a non-trivial
surface of two-dimensional gauge fields.

The same property holds for configurations with higher number of
instantons.
 For instance, the  gauge field configuration with two instantons
  symmetrically centered at
$x_+=(x_0,0,0,0)$ and $x_-=(-x_0,0,0,0)$ and one single scale $\rho$
reads
 $$ A_\mu= \overline{\tau}_{\mu \nu} \partial^\nu \phi(x) $$
with 
$$\phi(x)=\log\left(1 +{\rho^2\over (x-x_+)^2}+{\rho^2\over 
(x-x_-)^2}\right) $$ 
and $\overline{\tau}_{\mu\nu}= i(
\overline{\tau}_\nu^{\dagger}  \overline{\tau}_\mu - \delta_{\mu\nu})/2$,
$\overline{\tau}=(-I,-i\vec{\sigma})$.

 The  corresponding map $[\j^\ast A]:S^{2}\longrightarrow\CM^{S^{2}}$  induced
by the maximal holomorphic gauge  also reaches the boundary of the
maximal domain $\CM^{S^2}\back\CM^{S^2}_0$ because
the two-dimensional gauge field $\j^\ast_{(0,0)} A$ is again the abelian Dirac
monopole \mm\ with unit magnetic charge in
$\CM^{S^2}\back\CM^{S^2}_0$ \unst. 

In fact it can be shown that this property is   satisfied by
any gauge field carrying a non-trivial topological charge.
 This is not surprising because there are
topological reasons which imply that
 generic  gauge fields with multi-instantons  must induce  at least
one non-abelian monopole in some 2-dimensional spheres of
$S^4$\unst.  More precisely,
  a generic (non-symmetric) SU(N) gauge field $A$ 
 with non-trivial second Chern class
 $c_2(A)=k$ induces  a map $[\j^\ast
A]:S^{2}\longrightarrow\CM^{S^{2}}$ which  belongs to the $k$ 
 class  of the second homotopy group of $\CM$, $\pi_2(\CM)=\IZ $
 \asomitt.  Then, since the fundamental domain $\CM^{S^2}_0$
is contractible and, thus, homotopically trivial the image of
$[\j^\ast A]$ cannot be completely contained in $\CM^{S^2}_0$ \foot{
The result is to some extent  dual of the descendent technique in
the study of anomalies \ref\stora{R. Stora, in {\it Progress
in Gauge Field Theory}, Carg\`ese Lectures 1983, Eds. G. 't Hooft et {\it al},
Plenum Press, NY (1984)}}.

However, the connection with non-abelian monopoles is not exclusive
of 4-dimensional gauge fields with non-trivial topological charge.
As we have seen in previous sections there is a codimension zero
sector of 4-dimensional gauge fields with trivial topological charge
whose induced maps $[\j^\ast
A]:S^{2}\longrightarrow \CM^{S^{2}}$ do not lie
 inside $\CM^{S^{2}}_0$. 

All these facts suggests that 
the obstruction to the extension of the domain of the non-abelian
gauges described in this paper is based in physical grounds. We know
by general topological arguments that a complete global gauge cannot
exist but the special characteristics of the maximal non-abelian
gauges  point out in an intrinsic way which configurations
are relevant for some low energy non-perturbative effects. The
characterization of those configurations in terms of non-abelian 
magnetic monopoles provides a sound basis for a  physical
realization of the 't Hooft-Mandelstam confinement mechanism.

\newsec{Conclusions}

In infinite volume we can consider a similar construction
based on a family of $S^{d-1}$ spheres centered at the origin
of $\IR^d$ and with radius R varying from R$=0$ to R$=\infty$. 
Finiteness of Yang-Mills potential implies that  any
gauge configuration A  with finite energy verifies that $j^\ast_R A$ 
is a pure gauge field for  R$=0$ and R$=\infty$, 
which allow to associate to any field configuration $A$ with finite
potential energy  on $\IR^d$ one loop of gauge fields on $S^{d-1}$.
Iterating this procedure as in Sections 4 and 6 leads to the construction
of complete gauges for $d\leq 3$.

On the other hand the monopole identification also works for
field configurations in infinite volumes, which gives an intrinsic
physical meaning to the whole construction beyond the infrared limit.

In summary, the  two new gauge conditions introduced in Sections
4 and 6 based on the
special structure of the orbit spaces one and two-dimensional
gauge fields are complete for gauge fields over
spaces with dimensions lower than four in a maximal domain
$\CM_0$ which is open and dense in whole orbit space $\CM$.
The gauge conditions are free of Gribov ambiguities on $\CM_0$.

The obstruction to  completeness in four-dimensional spaces
is related to configurations describing a
dense gas or liquid of intantons which seem to be relevant for
confinement.

One of the interesting features of these maximal non-abelian
gauge conditions in lower dimensions is that the configurations
sitting at the boundary of the maximal domain do play  a very
relevant role in non-perturbative physical effects like the 
existence of nodes in the vacuum functional of 
Topologically Massive Gauge Theory  in 2+1 dimensions
and  Yang-Mills theory with 
$\theta=\pi$ in 3+1 dimensions. 
The disappearence of these nodes when the topological
mass or the $\theta$ term vanish suggest that those configurations
 play a fundamental role in the confinement mechanism which
is also activated when those terms vanish.

This characterization of gauge fields
over spaces of arbitrary dimension in terms of one or two dimensional
gauge fields provides an alternative description of the topological
properties of the corresponding orbit space. Homotopy and cohomology
classes of the orbit spaces $\CM^{S^d}$ are faithfully described in
terms of those of lower dimensional gauge fields, $\CM^{S^2}$ and
$\CM^{S^1}$. This description also makes possible an intrinsic
characterization of non-abelian monopoles for SU(N) gauge fields in terms
of field configurations lying beyond the boundary of the
fundamental domain of the maximal abelian gauges, which provides an
appropriate geometric framework for the realization of the
dual superconducting scenario for confinement. In fact,
the characterization of  gauge fields in terms of  SU(N)-
valued fields is closely related to the $\sigma$--model
chiral description QCD at low energies. This opens a
new perspective to the description of the
low energy glueball spectrum of pure gauge theories as an effective
theory described by an SU(N) sigma model.

\bigbreak\bigskip\bigskip\centerline {{\bf Acknowledgements}}
\nobreak  I thank  Carlo Becchi, Fernando Falceto, Pronob Mitter,
Mariano Santander and Andreas Wipf  
for valuable discussions on gauge fixing problems.
The work was completed at Erwing Schr\"odinger Institute.  I thank R. A.
Bertlmann for his hospitality. This  research is partially supported
by CICYT   grant AEN97-1680.  
\listrefs 

 \end